\documentclass[prb,twocolumn,showpacs,preprintnumbers,amssymb,superscriptaddress]{revtex4}
\usepackage{graphicx}% Include figure files
\usepackage[tbtags]{amsmath}
\usepackage{bm}% bold math

\begin{document}

\title
{Quantum theory of the low-frequency linear susceptibility of
interferometer-type superconducting qubits}
\author{Ya. S. Greenberg}
\affiliation{Novosibirsk State Technical University, 20 Karl Marx
Avenue, 630092 Novosibirsk, Russia}
\author{E. Il'ichev}
\affiliation{Institut f\"ur Photonische Technologien, Jena,
Germany}

\date{\today}
\begin{abstract}
We use the density matrix formalism to analyze the interaction of
interferometer-type superconducting qubits with a high quality
tank circuit, which frequency is well below the gap frequency of a
qubit. We start with the ground state characterization of the
superconducting flux and charge qubits. Then, by making use of a
dressed state approach we describe the qubits' spectroscopy when
the qubit is irradiated by a microwave field which is tuned to the
gap frequency. The last section of the paper is devoted to
continuous monitoring of qubit states by using a DC SQUID in the
inductive mode.

\end{abstract}

\maketitle
\section{Introduction}

The existence of a superposition between macroscopically distinct
quantum states has important implications, in particular, for the
quantum measurement theory, since a single macroscopic quantum
system can be easily accessed by the macroscopic measuring device.
One of the most advanced solid state quantum systems is a
superconducting qubit which is based on either the charge or phase
degrees of freedom\cite{Shnirman, Wendin}. Several ways of reading
out the qubit properties have been proposed and implemented. In an
initial research stage switching current measurements combined
with the excitation of a qubit by microwave pulses were
used\cite{Wal, Nisk}. Recently it turns out that in some cases
inductive (dispersive) measurements can be more effective. In the
frame of this approach the qubit is coupled with
linear\cite{Green3, Graj4, wal, Blais, Zorin, krec1, Born, Shnyr,
Zang} or non-linear oscillators\cite{Lup, Lup1, sid}.

In this paper we study two kinds of interferometer type
superconducting qubits (flux and charge qubits) which are realized
in practice as Josephson junctions embedded in a superconducting
(interferometric) loop\cite{Wal, Zorin, Born}.

A distinct feature of our approach is that a qubit is inductively
coupled with a high quality tank circuit (a linear $L_{T}C_{T}$
oscillator, where $L_{T}$ is the inductance and $C_{T}$ is the
capacitance of the tank) which resonance frequency is well below
the gap frequency of the qubit. The essential information about
the qubit properties can be extracted from the voltage across the
tank. The method is found to be a reliable technique for the
investigation of the properties of the flux qubit
systems\cite{Ploeg2, Izmal, Graj1, Graj2}, (prior to 2004 see the
review paper\cite{Ilich} and references therein).

Our approach here is based on the rigorous quantum mechanical
formulation of the problem, which allows us to apply it to the
ground state characterization not only of the flux
qubit\cite{Green3}, but of the charge qubit as well. In addition,
we show in Section 3 that our method is extremely useful for the
investigation of the spectroscopic properties of superconducting
qubits.

Since the qubit characteristic frequencies are well above the tank
resonance we may consider the tank circuit as a classical system
while the qubit is treated quantum mechanically. This description
has been proposed in Refs.~\onlinecite{Green3},
\onlinecite{Green4}, and the present work is its natural
generalization and continuation\cite{com3}.

Therefore, the voltage, $V$, across the tank obeys the equation:
\begin{equation}\label{volt}
\ddot V + \gamma _T \dot V + \omega _T^2 V = -M\omega _T^2
\frac{{d\langle\widehat{I}_q \rangle}}{{dt}} + \omega _T^2 L_T
\dot I_b,
\end{equation}
where  $\gamma_T= \omega_T/Q_T$, $Q_T$ is the quality factor of
the unloaded tank, $M$ is the mutual inductance between the tank
and the qubit loop, $\omega_T=(L_TC_T)^{-1/2}$ is the tank
resonance frequency, and $I_b(t)=I_0\cos\omega t$ is the tank bias
current with bias frequency $\omega$. Here $\langle
\widehat{I_q}\rangle$ is the quantum mechanical average of the
quantum operator of the current $\widehat{I_q}$ in the
interferometer loop.

Eq. (\ref{volt}) in Fourier components reads:
\begin{equation}\label{Vomega}
  V(\omega )\left( {\omega _T^2  - \omega ^2  + i\omega \gamma _T } \right) =
  -M\omega _T^2 \left(\frac{{d\langle\widehat{I}_q
  \rangle}}{{dt}}\right)_{\omega}
  +i\omega \omega _T^2 L_TI_0.
\end{equation}
 In what follows we assume that the interaction between
the qubit and the tank is weak. In this case it would be
reasonable to assume (below we prove this assumption by detailed
calculations) that
\begin{equation}\label{Curr_drv}
    \left(\frac{{d\langle\widehat{I}_q
    \rangle}}{{dt}}\right)_{\omega}=MI_T(\omega)Z(\omega),
\end{equation}
where $Z(\omega)$ is a complex function
($Z(\omega)=Z_1(\omega)+iZ_2(\omega)$) which depends on the qubit
properties only. As is seen from (\ref{Curr_drv}) the quantity
${{d\langle\widehat{I}_q \rangle}}/{{dt}}$ is different from zero
only due to the interaction of the qubit with the tank.

By taking into account that $I_T(\omega)=-iV(\omega)/\omega L_T$,
we obtain for the tank detuning $\xi$, and friction $\Gamma_T$:
\begin{equation}\label{ksi}
    \xi=\omega_T^2-\omega^2+\frac{k^2L_q\omega_T^2}{\omega}Z_2(\omega),
\end{equation}
\begin{equation}\label{GammaT}
    \Gamma_T=\gamma_T-\frac{k^2L_q\omega_T^2}{\omega^2}Z_1(\omega),
\end{equation}
where $L_q$ is the inductance of the qubit loop, $k^2=M^2/L_TL_q$
is a coupling constant, which characterizes the inductive
interaction between the qubit and the tank.

From ~(\ref{ksi}) and ~(\ref{GammaT}) we obtain the voltage
amplitude $V_T$, and the phase $\alpha$: $V_T  = \omega \omega
_T^2 L_T I_0 /\sqrt {\xi ^2  + \omega ^2 \Gamma _T^2 }$, and
$\tan\alpha = \xi /\omega \Gamma _T$.

Below we calculate in a rigorous quantum mechanical way the low
frequency linear response of a qubit to a low frequency probe
signal. It is necessary to say that our approach is valid if the
interaction between the qubit and the tank is weak which allows us
to neglect all nonlinear terms caused by the finite amplitude of
the low-frequency signal. Therefore, the main aim here is to
calculate the quantity $Z(\omega)$ which transfers the qubit
properties to the low frequency characteristics of the tank
circuit: its detuning (\ref{ksi}) and friction (\ref{GammaT}).

The paper is organized as follows. In Section 2 the general
approach to the ground state characterization of qubits is
considered. At the end of the section we obtain the expressions
for detuning and friction both for the flux and charge qubits. In
Section 3 we consider an interferometer type qubit which is
inductively coupled to a low frequency resonant circuit and
subjected simultaneously to microwave radiation. By making use of
the dressed state approach\cite{Coen} we calculate the low
frequency susceptibilities for charge and flux qubits. The results
obtained in this section are applied to explain the recently found
phenomenon of the damping and amplification of the output signal
by a superconducting flux qubit\cite{Graj}. The method developed
in this paper is applied in Section 4 to a continuous readout of a
superconducting qubit by a DC SQUID in the inductive mode.

\section{Ground state characterization of qubits}

In this section we consider a qubit which is isolated from
microwave fields and interacts inductively with a high quality
tank circuit. In what follows we assume the tank frequency is well
below the qubit gap frequency $\Delta E$ and, in addition, the
temperature is sufficiently low, so that $k_BT<<\Delta E$. Under
these conditions the signal detected at the output of the tank
circuit is mainly defined by the properties of the qubit ground
state.

\subsection{Rate equation for two level system}

In the basis of eigenstates $|1\rangle$, $|2\rangle$ the
Hamiltonian of a qubit reads:

\begin{equation}\label{Ham2}
H_0 = \frac{\Delta E}{2}\sigma _Z,
\end{equation}
where $\Delta E$ is the gap between two energy states. The
eigenstates are denoted in the following as $|1\rangle$ (ground
state) and $|2\rangle$ (excited state) with the properties:
$\sigma_Z|1\rangle=-|1\rangle$, $\sigma_Z|2\rangle=|2\rangle$,
$\sigma_X|1\rangle=|2\rangle$, $\sigma_X|2\rangle=|1\rangle$.
$\sigma_Y|1\rangle=i|2\rangle$, $\sigma_Y|2\rangle=-i|1\rangle$.

Next we introduce the density matrix:
\begin{equation}\label{dm}
    \frac{d\sigma}{dt}=-\frac{i}{\hbar}[H_0,\sigma],
\end{equation}
and define its matrix elements as
$\rho_{11}=\langle1|\sigma|1\rangle$,
$\rho_{22}=\langle2|\sigma|2\rangle$,
$\rho_{12}=\langle1|\sigma|2\rangle$,
$\rho_{21}=\langle2|\sigma|1\rangle=\rho_{12}^+$. From (\ref{dm})
we find the equations for the elements of the density matrix:
\begin{equation}\label{dm11}
    \frac{d\rho_{11}}{dt}=0,
\end{equation}
\begin{equation}\label{dm22}
    \frac{d\rho_{22}}{dt}=0,
\end{equation}
\begin{equation}\label{dm12}
    \frac{d\rho_{12}}{dt}=-i\Omega\rho_{12},
\end{equation}
where $\Omega=\Delta E/\hbar$. In the case of damping the above
equations should be changed to:
\begin{equation}\label{dm11damp}
    \frac{d\rho_{11}}{dt}=-\Gamma_\uparrow\rho_{11}+\Gamma_\downarrow\rho_{22},
\end{equation}
\begin{equation}\label{dm22damp}
    \frac{d\rho_{22}}{dt}=\Gamma_\uparrow\rho_{11}-\Gamma_\downarrow\rho_{22},
\end{equation}
\begin{equation}\label{dm12damp}
    \frac{d\rho_{12}}{dt}=-i\Omega\rho_{12}-\Gamma_\varphi\rho_{12},
\end{equation}
where $\Gamma_{\downarrow}$ is the transition rate from state
$|2\rangle$ to state $|1\rangle$ (relaxation rate),
$\Gamma_{\uparrow}$ is the transition rate from state $|1\rangle$
to state $|2\rangle$ (excitation rate), and the quantity
$\Gamma_{\varphi}$ is the rate of decoherence. For equilibrium
conditions the relaxation and excitation rates are related by the
detailed balance law:
\begin{equation}\label{balance}
\Gamma_{\uparrow}=\Gamma_{\downarrow}\exp\left(-\frac{\Delta
E}{k_BT}\right).
\end{equation}
From (\ref{balance}) we obtain
\begin{equation}\label{balance1}
   \Gamma_-T_1\equiv-\rho^{(eq)}=-\tanh\left(\frac{\Delta
   E}{2k_BT}\right),
\end{equation}
where we define $\Gamma_-=\Gamma_{\uparrow}-\Gamma_{\downarrow}$
and the longitudinal relaxation time
$T^{-1}_1=\Gamma_{\uparrow}+\Gamma_{\downarrow}$.

We rewrite Eqs. (\ref{dm11damp}), (\ref{dm22damp}),
(\ref{dm12damp}) in operator form
\begin{equation}\label{dm1}
    \frac{d\sigma}{dt}=-\frac{i}{\hbar}[H_0,\sigma]+\widehat{L},
\end{equation}
where
\begin{multline}\label{L}
\widehat{L}=-\Gamma_\uparrow|1\rangle\langle1|\sigma|1\rangle\langle1|
+\Gamma_\downarrow|1\rangle\langle2|\sigma|2\rangle\langle1|\\
+\Gamma_\uparrow|2\rangle\langle1|\sigma|1\rangle\langle2|
-\Gamma_\downarrow|2\rangle\langle2|\sigma|2\rangle\langle2|\\
-\Gamma_\varphi|1\rangle\langle1|\sigma|2\rangle\langle2|-
\Gamma_\varphi|2\rangle\langle2|\sigma|1\rangle\langle1|.
\end{multline}
As it is seen from (\ref{dm11damp}) and (\ref{dm22damp}), the
total population is constant:
$\frac{d}{dt}(\rho_{11}+\rho_{22})=0$. We take the normalization
condition as $\rho_{11}+\rho_{22}=1$. The rate equations can be
further simplified by introducing new variables:
$\rho=\rho_{11}-\rho_{22}$, which is the difference in populations
between the lower and the higher levels, and
$\rho_+=\rho_{12}+\rho_{21}$, $\rho_-=\rho_{12}-\rho_{21}$:
\begin{equation}\label{rho}
    \frac{d\rho}{dt}=-\frac{1}{T_1}\rho-\Gamma_-,
\end{equation}
\begin{equation}\label{rho-1}
    \frac{d\rho_{-}}{dt}=-i\Omega\rho_{+}-\Gamma_\varphi\rho_{-},
\end{equation}
\begin{equation}\label{rho+1}
    \frac{d\rho_{+}}{dt}=-i\Omega\rho_{-}-\Gamma_\varphi\rho_{+}.
\end{equation}
The quantity $\rho^{(eq)}$ in (\ref{balance1}) is just a steady
state solution of the equation (\ref{rho}), and is the difference
in equilibrium populations between the lower and the higher
levels.

\subsection{Interaction between qubit and tank circuit}

The Hamiltonian of a qubit which is coupled to the tank in the
eigenbasis reads:
\begin{equation}\label{Ham3}
H = \frac{\Delta E(\Phi_X)}{2}\sigma _Z +MI_T\widehat{I}_q,
\end{equation}
where $I_T$ is the current in the tank inductance, $\Phi_X$ is a
DC bias flux through a qubit loop.

The current  operator of the qubit, $\widehat{I}_q$,
 can generally be written as:
\begin{equation}\label{Curr}
    \widehat{I}_q=I_X\sigma_X+I_Y\sigma_Y+I_Z\sigma_Z,
\end{equation}
where the quantities $I_X$, $I_Y$, $I_Z$, which will be specified
below, depend on the nature of the qubit.

The interaction with the tank also influences the $\Gamma$ rates
in Eqs. (\ref{dm11damp}), (\ref{dm22damp}). Therefore, we may
write in linear approximation
\begin{equation}\label{gamma_up_dir}
    \Gamma_{\uparrow}^{(\lambda)}=\Gamma_{\uparrow}+\lambda
    \frac{d\Gamma_{\uparrow}}{d\phi_X},
\end{equation}
\begin{equation}\label{gamma_dn_dir}
    \Gamma_{\downarrow}^{(\lambda)}=\Gamma_{\downarrow}+\lambda
    \frac{d\Gamma_{\downarrow}}{d\phi_X},
\end{equation}
where $\lambda=2\pi MI_T/\Phi_0$, $\phi_X=2\pi\Phi_X/\Phi_0$,
$\Phi_0=h/2e$ is the flux quantum.

As for $\Gamma_{\varphi}$, we assume this rate is sufficiently
high which allows us to neglect its modification by slow time
dependent external flux $MI_T$.

The equation for the density matrix is similar to (\ref{dm1})
\begin{equation}\label{dm2}
    \frac{d\sigma}{dt}=-\frac{i}{\hbar}[H,\sigma]+\widehat{L},
\end{equation}
where $H$ is given in (\ref{Ham3}).

From (\ref{dm2}) we get the following equations for the elements
of the density matrix:
\begin{equation}\label{ro}
    \frac{d\rho}{dt}=\frac{2i\lambda I_X}{\hbar}
    \frac{\Phi_0}{2\pi}\rho_--
    \frac{2\lambda I_Y}{\hbar}\frac{\Phi_0}{2\pi}\rho_+-\frac{1}{T_1}\rho-\Gamma_-
    +\lambda\frac{1}{T_1}\frac{d\rho^{(eq)}}{d\phi_X},
\end{equation}
\begin{equation}\label{ro-}
    \frac{d\rho_{-}}{dt}=i\Omega\rho_{+}-\Gamma_\varphi\rho_-
    +\frac{2i\lambda I_Z}{\hbar}\frac{\Phi_0}{2\pi}\rho_+
    +\frac{2i\lambda I_X}{\hbar}\frac{\Phi_0}{2\pi}\rho,
\end{equation}
\begin{equation}\label{ro+}
    \frac{d\rho_{+}}{dt}=i\Omega\rho_{-}-\Gamma_\varphi\rho_{+}
    +\frac{2i\lambda I_Z}{\hbar}\frac{\Phi_0}{2\pi}\rho_{-}
    +\frac{2\lambda I_Y}{\hbar}\frac{\Phi_0}{2\pi}\rho.
\end{equation}

\subsection{Linear susceptibilities for the qubit}
We find the solution to Eqs. (\ref{ro}) - (\ref{ro-}), by assuming
the coupling $\lambda$ is small. Therefore, the time dependent
solution for these equations can be obtained by the perturbation
method as the small time dependent corrections to the steady state
values: $\rho(t)=\rho^{(0)}+\rho^{(1)}(t)$,
$\rho_+(t)=\rho_+^{(0)}+\rho^{(1)}_+(t)$ ,
$\rho_-(t)=\rho_-^{(0)}+\rho^{(1)}_-(t)$, where $\rho^{(0)}$,
$\rho^{(0)}_+$, and $\rho^{(0)}_-$ are steady state solutions for
Eqs. (\ref{ro}) - (\ref{ro-}) with $\lambda=0$. For these steady
state values we readily obtain: $\rho^{(0)}=\rho^{(eq)}$,
$\rho^{(0)}_+=0$, $\rho^{(0)}_-=0$. Therefore, the fist order
corrections $\rho^{(1)}(t)$, $\rho^{(1)}_+(t)$, $\rho^{(1)}_-(t)$
satisfy the following equations (below we omit superscript 1):
\begin{equation}\label{ro1}
    \frac{d\rho}{dt}=-\frac{1}{T_1}\rho+\lambda\frac{1}{T_1}\frac{d\rho^{(eq)}}{d\phi_X},
\end{equation}
\begin{equation}\label{ro-1}
    \frac{d\rho_{-}}{dt}=i\Omega\rho_{+}-\Gamma_\varphi\rho_{-}
    +\frac{2i\lambda I_X}{\hbar}\frac{\Phi_0}{2\pi}\rho^{(eq)},
\end{equation}
\begin{equation}\label{ro+1}
    \frac{d\rho_{+}}{dt}=i\Omega\rho_{-}-\Gamma_\varphi\rho_{+}
    +\frac{2\lambda I_Y}{\hbar}\frac{\Phi_0}{2\pi}\rho^{(eq)}.
\end{equation}

 From these equations it is not difficult to find the linear
susceptibilities of the system
($\chi_{\rho}(\omega)=\rho(\omega)/\lambda(\omega)$, etc., where
$\lambda(\omega)=2\pi MI_T(\omega)/\Phi_0$). Therefore, we get
\begin{equation}\label{hiro}
    \chi_{\rho}(\omega)=\frac{1}{(i\omega
    T_1+1)}\frac{d\rho^{(eq)}}{d\phi_X},
\end{equation}

\begin{equation}\label{hiro-}
    \chi_{\rho_-}(\omega)=\frac{2i\Phi_0\rho^{(eq)}}{2\pi\hbar d(\omega)}
    \left[\left(i\omega+\Gamma_{\varphi}\right)I_X+\Omega
    I_Y\right],
\end{equation}
\begin{equation}\label{hiro+}
 \chi_{\rho_+}(\omega)=\frac{2\Phi_0\rho^{(eq)}}{2\pi\hbar d(\omega)}
    \left[\left(i\omega+\Gamma_{\varphi}\right)I_Y-\Omega
    I_X\right],
\end{equation}
where
\begin{equation}\label{domega}
    d(\omega)=\left(i\omega+\Gamma_{\varphi}\right)^2+\Omega^2.
\end{equation}
\subsection{Calculation of $Z_1(\omega)$ and $Z_2(\omega)$}
First we calculate the average current
$\langle\widehat{I}_q\rangle$. By using (\ref{Curr}) we obtain:

\begin{equation}\label{curr1}
    \langle\widehat{I}_q\rangle=I_XTr(\sigma\sigma_X)+I_YTr(\sigma\sigma_Y)+I_ZTr(\sigma\sigma_Z),
\end{equation}
where $Tr(\sigma\sigma_Z)=-\rho$,
$Tr(\sigma\sigma_X)=\rho_+$,
$Tr(\sigma\sigma_Y)=i\rho_-$. Therefore,
\begin{equation}\label{Curr1}
\langle \widehat{I}_q\rangle  = -I_Z\rho+I_X\rho_++iI_Y\rho_-.
\end{equation}
With the aid of equations (\ref{ro}), (\ref{ro-}), and (\ref{ro+})
we obtain:
\begin{multline}\label{Curr2a}
\frac{d\langle
\widehat{I}_q\rangle}{dt}=\frac{I_Z}{T_1}\left(\rho+\Gamma_-T_1-\lambda\frac{d\rho^{(eq)}}{d\phi_X}\right)\\
+i\left(I_X\Omega-I_Y\Gamma_{\varphi}\right)\rho_-
-\left(I_X\Gamma_{\varphi}+I_Y\Omega\right)\rho_+.
\end{multline}
If the interaction between the qubit and the tank is absent
($\lambda=0$), then, $\rho^{(0)}=\rho^{(eq)}$, $\rho^{(0)}_+=0$,
$\rho^{(0)}_-=0$, and, as is seen from (\ref{Curr1}) and
(\ref{Curr2a}), the average current is proportional to the
difference of equilibrium populations  between the two states:
$\langle \widehat{I}_q\rangle =-I_Z\rho^{(eq)}$, but
${d\langle\widehat{I}_q\rangle}/{dt}=0$. However, if the
interaction is sufficiently weak, such that $\lambda \ll 1$, the
first approximation for the latter quantity gives:
\begin{multline}\label{Curr3}
\frac{d\langle
\widehat{I}_q\rangle}{dt}=\frac{I_Z}{T_1}\left(\rho^{(1)}-\lambda\frac{d\rho^{(eq)}}{d\phi_X}\right)\\
+i\left(I_X\Omega-I_Y\Gamma_{\varphi}\right)\rho_-^{(1)}
-\left(I_X\Gamma_{\varphi}+I_Y\Omega\right)\rho_+^{(1)},
\end{multline}
and for its Fourier component:
\begin{multline}\label{Curr3a}
\left(\frac{d\langle \widehat{I}_q\rangle}{dt}\right)_\omega=
\frac{I_Z}{T_1}\left(\chi_{\rho}(\omega)-\frac{d\rho^{(eq)}}{d\phi_X}\right)\frac{2\pi
M}{\Phi_0}I_T(\omega)\\
+\frac{2\pi
M}{\Phi_0}I_T(\omega)\left[i\left(I_X\Omega-I_Y\Gamma_{\varphi}\right)\chi_{\rho_-}(\omega)\right.\\
-\left.\left(I_X\Gamma_{\varphi}+I_Y\Omega\right)\chi_{\rho_+}(\omega)\right].
\end{multline}
 Therefore, we obtain for $Z_1(\omega)$ and $Z_2(\omega)$ defined in (\ref{Curr_drv}):
 \begin{multline}\label{z1}
    Z_1(\omega)=\frac{2\pi}{\Phi_0}\frac{I_Z}{T_1}\left(\chi'_{\rho}(\omega)-\frac{d\rho^{(eq)}}{d\phi_X}\right)\\
    -\frac{2\pi}{\Phi_0}\left(I_X\Omega-I_Y\Gamma_{\varphi}\right)\chi''_{\rho_-}(\omega)
    -\frac{2\pi}{\Phi_0}\left(I_X\Gamma_{\varphi}+I_Y\Omega\right)\chi'_{\rho_+}(\omega),
 \end{multline}
 \begin{multline}\label{z2}
    Z_2(\omega)=\frac{2\pi}{\Phi_0}\frac{I_Z}{T_1}\chi''_{\rho}(\omega)\\
    +\frac{2\pi}{\Phi_0}\left(I_X\Omega-I_Y\Gamma_{\varphi}\right)\chi'_{\rho_-}(\omega)
    -\frac{2\pi}{\Phi_0}\left(I_X\Gamma_{\varphi}+I_Y\Omega\right)\chi''_{\rho_+}(\omega),
 \end{multline}
 where $\chi'_{\rho}(\omega)$, $\chi'_{\rho_-}(\omega)$, $\chi'_{\rho_+}(\omega)$ and
 $\chi''_{\rho}(\omega)$, $\chi''_{\rho_-}(\omega)$, $\chi''_{\rho_+}(\omega)$
 are, respectively, real and imaginary parts of the corresponding susceptibilities
 (\ref{hiro}) - (\ref{hiro+}).

If we account for the fact that the gap frequency is large,
$\Omega\gg\omega, \Gamma_{\varphi}$, we can simplify the
calculations  of $Z_1(\omega)$ and $Z_2(\omega)$ to obtain the
tank detuning $\xi$, and friction, $\Gamma_T$ in the following
form:

\begin{multline}\label{ksi chrg}
    \xi=\omega_T^2-\omega^2 -I_Z\frac{k^2\omega_T^2 L_q}{
    \left(1+\omega^2 T_1^2\right)}\frac{d\rho^{(eq)}}{d\Phi_X}
    \\-\frac{2k^2L_q\omega_T^2}{\Delta
    E}\rho^{(eq)}\left(I_X^2+I_Y^2\right),
    \end{multline}
\begin{multline}\label{GammaT1_chrg}
    \Gamma_T=\gamma_T+I_Z\frac{k^2\omega_T^2 L_qT_1}{
    \left(1+\omega^2 T_1^2\right)}\frac{d\rho^{(eq)}}{d\Phi_X}\\+
    \frac{4k^2L_q\hbar^2\omega_T^2\Gamma_{\varphi}}{(\Delta
    E)^3}\rho^{(eq)}\left(I_X^2+I_Y^2\right).
\end{multline}
The expressions (\ref{ksi chrg}), (\ref{GammaT1_chrg}) are
applicable for any kind of interferometer type superconducting
qubit once the components $I_X$, $I_Y$, $I_Z$ of the current
operator are known.

The terms in right hand sides of (\ref{ksi chrg}) and
(\ref{GammaT1_chrg}) which are proportional to
${d\rho^{(eq)}}/{d\Phi_X}$  reflect the effect of thermalization
which value depends on the interplay between the relaxation rate
$T_1^{-1}$ and the tank frequency $\omega$.

The quantity ${d\rho^{(eq)}}/{d\Phi_X}$ can be expressed as:
\begin{equation}
    \frac{d\rho^{(eq)}}{d\Phi_X}=-\frac{\cosh^{-2}\left(\frac{\Delta
    E}{2k_BT}\right)}{k_BT}I_Z\nonumber,
\end{equation}
where we used the fact that $I_Z$ can be written  as the
derivative of the ground state energy $E_G$ over the magnetic
flux: $I_Z=dE_G/d\Phi_X$, where $E_G=-\Delta E/2$. Since for a
proper qubit operation the condition $\Delta E>>k_BT$ is
necessary, the quantity ${d\rho^{(eq)}}/{d\Phi_X}$ scales as
$\exp\left(-\frac{\Delta E}{k_BT}\right)$. Therefore, no matter
what the value $\omega T_1$ is, the contribution to the tank
response of the terms in (\ref{ksi chrg}) and (\ref{GammaT1_chrg})
which are proportional to ${d\rho^{(eq)}}/{d\Phi_X}$ can be
neglected as compared with the contribution of the last terms in
these expressions.

\subsection{Low frequency response of the flux qubit}
For the flux qubit the energy gap is
\begin{equation}\label{engap}
    \Delta
    E\equiv\Delta_{\varepsilon}=\sqrt{\Delta^2+\varepsilon^2},
\end{equation}
where $\Delta/2$ is the tunnelling amplitude between degenerate
flux states. The bias $\varepsilon$ is controlled by an external
dc flux $\Phi_X$: $\varepsilon=2\Phi_0I_qf_X$, where $I_q$ is the
critical current of the flux qubit, $f_X=(\Phi_X/\Phi_0-1/2)$.

The current operator of the flux qubit in the eigenstate basis is:
\begin{equation}\label{curr_op_flux}
   \widehat{I}_q  = \frac{I_q}{\Delta_{\varepsilon}}\left(\varepsilon\sigma _Z  - \Delta\sigma _X\right).
\end{equation}

Therefore, the components of the current operator for a flux qubit
are as follows:
\begin{eqnarray}\label{Curr_comp_flux}
    I_X=-I_q\Delta/\Delta_{\varepsilon},
    I_Z=I_q\varepsilon/\Delta_{\varepsilon},
    I_Y=0.
\end{eqnarray}
From (\ref{ksi chrg}) and (\ref{GammaT1_chrg}) we obtain the
following expressions for the frequency detuning and the friction
of the flux qubit
\begin{equation}\label{ksi_flux}
    \xi=\omega_T^2-\omega^2
-\frac{2k^2\omega_T^2L_qI_q^2\Delta^2}{\Delta^3_{\varepsilon}}\rho^{(eq)},
\end{equation}
\begin{equation}\label{GammaT_flux}
    \Gamma_T=\gamma_T
+\frac{4k^2L_qI_q^2\Delta^2}{\Delta^3_{\varepsilon}}
    \left(\frac{\hbar\omega_T}{\Delta_{\varepsilon}}\right)^2\Gamma_{\varphi}\rho^{(eq)},
\end{equation}
where we neglected the terms which are proportional to
${d\rho^{(eq)}}/{d\Phi_X}$.

The last term in right hand side of (\ref{ksi_flux}), which gives
the main contribution of the  flux qubit to the tank detuning,
coincides with the result obtained earlier theoretically
\cite{Green3} and confirmed by experiment \cite{Graj4}.

The last term in (\ref{ksi_flux}) can be expressed in terms of the
curvature of the ground state\cite{Green3,Graj4}.
\begin{equation}\label{ksi_flux_main}
-\frac{2k^2\omega_T^2L_qI_q^2\Delta^2}{\Delta^3_{\varepsilon}}
\rho^{(eq)}=k^2\omega_T^2L_q\frac{d^2 E_G}{d\Phi_X^2}\rho^{(eq)}.
\end{equation}
This property is a direct consequence of the relation between
current components (\ref{Curr_comp_flux})
\[
\frac{dI_Z}{d\Phi_X}=2I_X^2,
\]
which seems to be peculiar for the flux qubit only. For other
types of superconducting qubits there is no simple relation
between the tank detuning and the curvature of the ground state.

As is seen from (\ref{GammaT_flux}), our theory predicts the
modification of the tank quality factor due to the interaction
with the qubit. In principle, this effect allows us to measure the
dephasing rate $\Gamma_{\varphi}$. However, for the parameters of
the flux qubit used in Ref.~\onlinecite{Graj4} this effect was
rather weak to be experimentally detected.

\subsection{Low frequency response of the charge qubit}

For a charge qubit with different critical currents, $I_{C1}$ and
$I_{C2}$ of its Josephson junctions the energy gap is

\begin{equation}\label{En_ch_qb}
    \Delta E=\sqrt{\epsilon^2_J+C^2},
\end{equation}
where $\epsilon^2_J=E_{J1}^2+E_{J2}^2+2E_{J1}E_{J2}\cos\phi_X$,
$E_{J1}=\Phi_0I_{C1}/2\pi$, $E_{J2}=\Phi_0I_{C2}/2\pi$,
$C=4E_C\left(1-n_g\right)$. Here the polarization charge on the
island $n_g$ is controlled by the gate voltage $V_g$ via the
capacitance $C_g$, namely  $n_g=C_{g}V_{g}/e$.

The components of the current operator are as follows\cite{Green}:
\begin{equation}\label{A}
I_Z=\frac{2\pi}{\Phi_0}\frac{E_{J1}E_{J2}}{2\Delta E}\sin\phi_X,
\end{equation}
\begin{multline}\label{B}
I_X=\frac{2\pi}{\Phi_0}\frac{E_{J1}+E_{J2}}{4\epsilon^2_J}\sin\frac{\phi_X}{2}\\
\times\left[\left(E_{J1}-E_{J2}\right)^2-\frac{4CE_{J1}E_{J2}}{\Delta
E}\cos^2\frac{\phi_X}{2}\right],
\end{multline}
\begin{multline}\label{D}
I_Y=\frac{2\pi}{\Phi_0}\frac{E_{J1}-E_{J2}}{4\epsilon^2_J}\cos\frac{\phi_X}{2}\\
\times\left[\left(E_{J1}+E_{J2}\right)^2+\frac{4CE_{J1}E_{J2}}{\Delta
E}\sin^2\frac{\phi_X}{2}\right].
\end{multline}
In case of symmetrical junctions ($E_{J1}=E_{J2}=E_J$) we have: $I_Y=0$,
\begin{equation}\label{IZ}
I_Z=\frac{\pi E_J}{\Phi_0\Delta E}\sin\phi_X,
\end{equation}
\begin{equation}\label{IX}
I_X=-\frac{\pi E_J C}{\Phi_0\Delta E}\sin\frac{\phi_X}{2},
\end{equation}
where
\begin{equation}\label{DE}
    \Delta E=\sqrt{4E_J^2\cos^2\frac{\phi_X}{2}+C^2}.
\end{equation}
For the tank detuning and friction we obtain from (\ref{ksi chrg})
and (\ref{GammaT1_chrg}), neglecting the terms which are
proportional to ${d\rho^{(eq)}}/{d\Phi_X}$:
\begin{equation}\label{ksi chrg1}
    \xi=\omega_T^2-\omega^2
    -\frac{2k^2L_q\omega_T^2}{\Delta
    E}\rho^{(eq)}\left(I_X^2+I_Y^2\right),
    \end{equation}
\begin{equation}\label{GammaT1_chrg1}
    \Gamma_T=\gamma_T+
    \frac{4k^2L_q\hbar^2\omega_T^2\Gamma_{\varphi}}{(\Delta
    E)^3}\rho^{(eq)}\left(I_X^2+I_Y^2\right).
\end{equation}
As distinct from the flux qubit, the tank detuning caused by the
charge qubit cannot, in general, be expressed in terms of the
ground state curvature. It can be done only for the symmetrical
case ($E_{J1}=E_{J2}$) at the point $\phi_X=\pi$.

From expression (\ref{ksi chrg1}) we calculate the dependence  of
the phase shift $\alpha$ of the output signal on the gate charge
parameter $n_g$ for different values of magnetic flux
$\Phi_X/\Phi_0$ applied to the qubit loop (see Fig.~\ref{phase
shift}). For the calculations we take the following values for the
tank, $Q_T=1000$, $k^2=0.01$, $\omega_T=2\pi\times 50$ MHz, and
for the charge qubit, $L_q=1.5$ nH, $E_{J1}/h=25$ GHz,
$E_{J2}/h=29$ GHz, $E_C/h=3.5$ GHz.
\begin{figure}
  \includegraphics[width=6 cm, angle=-90]{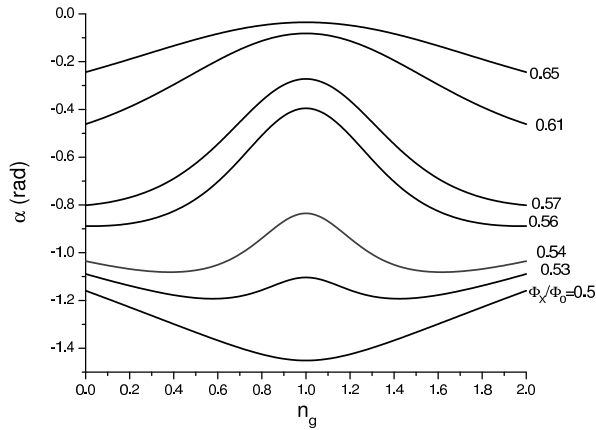}\\
\caption{The dependence of the phase shift $\alpha$ on the gate
parameter $n_g$ for different values of magnetic flux
$\Phi_X/\Phi_0$ applied to the qubit loop.}\label{phase shift}
\end{figure}

In conclusion to this section it is worth noting that the average
current in the qubit loop $\langle\widehat{I}_q\rangle$ is
proportional to the difference of equilibrium   populations
between the two qubit states only in the absence of interaction
between the qubit and the tank circuit: $\langle
\widehat{I}_q\rangle =-I_Z\rho^{(eq)}$. The other components of
the current, $I_X$ and $I_Y$ which give the main contribution to
the tank response (the last terms in (\ref{ksi chrg}) and
(\ref{GammaT1_chrg})), appear in the average current only due to
the interaction of the qubit with the tank.

\section{Characterization of irradiated qubit by Rabi spectroscopy}

The essence of the Rabi spectroscopy method is the following.
Under microwave irradiation, which frequency is close to the gap
frequency, the level structure of the global system (qubit,
radiation field and their interaction) is a ladder of pairs of
states where the spacing between two levels in the pair is equal
to the Rabi frequency, the value of which is controlled by the
power of microwaves\cite{Coen}. Normally, Rabi oscillations are
damped out with a rate, which depends on how strongly the system
is coupled to the environment. However, if a second low-frequency
source (in our case this source is a low frequency excitation from
the tank) is applied simultaneously to the qubit it responds with
persistent low frequency oscillations. The amplitude of these
low-frequency oscillations has a resonance at the Rabi frequency
and its width is dependent on the damping rates of the
system\cite{Green5, Green6}. Therefore, the interaction of the
qubit with the tank induces the transitions between the Rabi
levels which result in low frequency oscillations of the current
in a qubit loop which, in turn, result in the measurable response
of the tank\cite{Green1}. In particular, the signature of a high
frequency resonance can be read out from the low frequency
voltage-flux characteristic of the tank circuit.

Below we assume that the interaction between the qubit and
microwave  field does not influence the dephasing and the
relaxation rates. As was shown in Refs.~\onlinecite{Bloch} and
\onlinecite{Red}, this assumption is valid for relative weak
driving, sufficient short correlation time of the heat bath
$\tau_c$, and in the large temperature limit:
$F<<\Delta_{\varepsilon}, \hbar/\tau_c, k_BT$\cite{com}.

\subsection{Low frequency linear susceptibility of an irradiated qubit coupled to a tank circuit}

In this subsection we summarize in a concise form  the main
results from Ref.~\onlinecite{Green1}, which are relevant for
specific qubit applications.

The Hamiltonian for an irradiated qubit which interacts with a low
frequency tank circuit is as follows:

\begin{equation}\label{Ham3a}
H = \frac{\Delta E}{2}\sigma _Z
+\hbar\omega_0a^+a+H_{\texttt{int}}+H_{\texttt{int}}^{\texttt{LF}},
\end{equation}
where  the second term in (\ref{Ham3a}) describes a microwave
field, the third term describes the interaction of the qubit with
this microwave field:
\begin{equation}\label{H_int}
   H_{\texttt{int}} = - \frac{1}{2}\widehat{I}_q F(a^++a),
\end{equation}
where $\widehat{I}_q$ is the current operator of the qubit as
given in (\ref{Curr}),  $F$ is the amplitude of the microwave
field with the dimension of a magnetic flux. The last term in
(\ref{Ham3a}) describes the interaction of the qubit with a tank
circuit:
\begin{equation}\label{HamLF}
    H_{\texttt{int}}^{\texttt{LF}}=MI_T\widehat{I}_q=MI_T\left(I_Z\sigma_Z+I_X\sigma_X+I_Y\sigma_Y\right).
\end{equation}

We denote the eigenfunctions of the photon field as $|N\rangle$:
$a^+|N\rangle=\sqrt{N+1}|N+1\rangle$, and
$a|N\rangle=\sqrt{N}|N-1\rangle$. The eigenfunctions of the
noninteracting qubit and associated photon system we take in the
form of a tensor product
$|1,N\rangle\equiv|1\rangle\otimes|N\rangle$, and
$|2,N\rangle\equiv|2\rangle\otimes|N\rangle$.

If the photon frequency $\omega_0$ is close to the qubit gap
frequency $\Delta E/\hbar$, and the high frequency detuning
$\delta$ is small, $\delta=\omega_0-\Delta E/\hbar\ll \omega_0,
\Delta E/\hbar$, where for definitiveness we assume $\delta>0$,
then the energies of the states $|1,N+1\rangle$ and $|2,N\rangle$
are close to each other: $E_{1,N+1}-E_{2,N}=\hbar\delta$. The same
is true for the pairs of states $|1,N\rangle$ and $|2,N-1\rangle$,
$|1,N+2\rangle$ and $|2,N+1\rangle$, and so on. Therefore, the
energy levels of the system under consideration are a ladder of
pairs of states which are specified by the photon number $N$. The
spacing between two levels in the pair is equal to $\hbar\delta$,
and the distance between neighboring pairs is equal to the photon
energy $\hbar\omega_0$\cite{Coen}. These levels of uncoupled
qubit-photon system are modified due to the interaction
(\ref{H_int}). This interaction results in an increase of the
energy gap between two levels in the pair. The spacing between the
states $|1,N\rangle$ and $|2,N-1\rangle$ becomes equal to
$\hbar\Omega_R$, where $\Omega_R$ is the Rabi frequency
\cite{com1}
\begin{equation}\label{RF}
   \Omega_R=\sqrt{\delta^2+\Omega_1^2},
\end{equation}
where $\hbar\Omega_1=F\sqrt{I_X^2+I_Y^2}$.

From now on we will call these two nearby dressed
states Rabi levels.

As was shown in Ref.~\onlinecite{Green1} the interaction
(\ref{HamLF}) between the qubit and the tank results in the
transitions between Rabi levels. It is just these  transitions
which result in the low frequency response of a qubit detected by
the tank.

In Ref.~\onlinecite{Green1} we obtained the evolution equations
for the elements of the density matrix which describe the
transitions between these Rabi levels: $\rho$, $\rho_+$, and
$\rho_-$, where $\rho$ is the difference of the populations
between higher and lower Rabi levels. (Note, that here the
definition of $\rho$ is different from that given in Section II).
These elements of the density matrix are usually accounted for by
a so called rotating wave approximation (RWA). The equations for
the elements of the density matrix which describe in RWA the
interaction of an irradiated qubit with the tank are as follows:
\begin{equation}\label{roLF}
    \frac{d\rho}{dt}=-A_1\rho+B\rho_++
    \frac{2i\lambda\Phi_0}{2\pi\hbar}I_Z\sin2\theta\rho_-
   +\Gamma_-\cos2\theta,
\end{equation}

\begin{equation}\label{roLF2}
    \frac{d\rho_+}{dt}=-i\Omega_R\rho_-+B\rho-A_2\rho_+- \frac{2i\lambda\Phi_0}{2\pi\hbar}I_Z\cos2\theta\rho_-\\
   +\Gamma_-\sin2\theta,
   \end{equation}

\begin{multline}\label{roLF3}
     \frac{d\rho_-}{dt}=-i\Omega_R\rho_+-\Gamma_{\varphi}\rho_--\frac{2i\lambda\Phi_0}{2\pi\hbar}I_Z
     \left(\rho_+\cos2\theta-\rho\sin2\theta\right),
\end{multline}

\begin{equation}\label{A1}
    A_1=\left[\frac{1}{T_1}\cos^22\theta+\Gamma_{\varphi}\sin^22\theta\right],
\end{equation}
\begin{equation}\label{A2}
    A_2=\left[\frac{1}{T_1}\sin^22\theta+\Gamma_{\varphi}\cos^22\theta\right],
\end{equation}
\begin{equation}\label{B1}
    B=\left[\Gamma_{\varphi}-\frac{1}{T_1}\right]\sin2\theta\cos2\theta.
\end{equation}
The angle $\theta$ is given by $\tan2\theta=-\Omega_1/\delta$,
where $0<2\theta<\pi$, so that $\cos2\theta=-\delta/\Omega_R$,
$\cos\theta=\frac{1}{\sqrt{2}}\left(1-\frac{\delta}{\Omega_R}\right)^{1/2}$, and
$\sin\theta=\frac{1}{\sqrt{2}}\left(1+\frac{\delta}{\Omega_R}\right)^{1/2}$.

The steady state solution for the elements of the density matrix
and the low frequency linear susceptibility of a qubit are as
follows \cite{Green1}:
\begin{equation}\label{st0}
    \rho^{(0)}=\frac{\left(\Gamma_{\varphi}^2+\Omega_R^2\right)}
    {\frac{\Gamma_{\varphi}^2}{T_1}+A_1\Omega_R^2}\Gamma_-\cos2\theta,
\end{equation}

\begin{equation}\label{st+}
    \rho_+^{(0)}=\frac{\Gamma_{\varphi}^2}{\frac{\Gamma_{\varphi}^2}{T_1}+A_1\Omega_R^2}\Gamma_-\sin2\theta,
\end{equation}

\begin{equation}\label{st-}
    \rho_-^{(0)}=-i\frac{\Omega_R}{\Gamma_{\varphi}}\rho_+^{(0)},
\end{equation}

\begin{widetext}
\begin{equation}\label{hiroR}
    \chi_{\rho}(\omega)=\frac{2\Phi_0\Omega_R}{D(\omega)2\pi\hbar\Gamma_{\varphi}}I_Z
   \rho_+^{(0)}\left[\sin2\theta\left[\left(i\omega+
   \Gamma_{\varphi}\right)\left(i\omega+\frac{1}{T_1}\right)+\Omega^2_R\right]
   +\frac{\Omega_R^2}{\Gamma_{\varphi}}B\cos2\theta\right],
\end{equation}

\begin{equation}\label{hiro+R}
\chi_{\rho_+}(\omega)=-\frac{2\Phi_0\Omega_R}{D(\omega)2\pi\hbar\Gamma_{\varphi}}I_Z \rho_+^{(0)}\cos2\theta
 \left[\left(i\omega+\Gamma_{\varphi}\right)\left(i\omega+\frac{1}{T_1}\right)
  -\left(i\omega+A_1\right)\frac{\Omega_R^2}{\Gamma_{\varphi}}\right],
\end{equation}

\begin{equation}\label{hiro-R}
\chi_{\rho_-}(\omega)=i\frac{2\Phi_0\Omega_R^2}
{D(\omega)2\pi\hbar\Gamma_{\varphi}^2}I_Z\rho_+^{(0)}\cos2\theta
\left(i\omega+\frac{1}{T_1}\right)\left(i\omega+2\Gamma_{\varphi}\right),
\end{equation}
where
\begin{equation}\label{Domega}
    D(\omega)=\left(i\omega+\Gamma_{\varphi}\right)^2
    \left(i\omega+\frac{1}{T_1}\right)+\left(i\omega+A_1\right)\Omega_R^2.
\end{equation}
\end{widetext}

It is interesting to note that under high frequency irradiation
the population of the Rabi levels becomes inverted. It is seen
from (\ref{st0}), where the quantity $\rho^{(0)}$, which is, by
definition, the difference  of the populations between higher and
lower Rabi levels, is positive, since for $\delta>0$ we have
$\cos2\theta=-\delta/\Omega_R<0$, and always $\Gamma_-<0$. In
addition, as $\delta$ tends to zero, $\rho^{(0)}\rightarrow 0$
which causes the equalization of the population of the two levels
($\rho_{11}=\rho_{22}=\frac{1}{2}$) when the high frequency
irradiation is in exact resonance with the energy gap of the
qubit.

\subsection{Calculation of $Z_1(\omega)$ and $Z_2(\omega)$ for irradiated qubit}

Since the matrix elements of $\sigma_X$ and $\sigma_Y$ between the
Rabi levels are zero, the average current is:
\begin{equation}\label{curr1a}
    \langle\widehat{I}_q\rangle=I_ZTr(\sigma\sigma_Z)=I_Z\left(\rho(t)\cos2\theta+\rho_+(t)\sin2\theta\right).
\end{equation}
By using Eqs. (\ref{roLF}) - (\ref{roLF3}) we obtain:
\begin{multline}\label{dotcurr}
\frac{{d\langle\widehat{I}_q \rangle}}{{dt}}=-\frac{I_Z
}{T_1}\left[\rho(t)\cos2\theta+\rho_+(t)\sin2\theta\right.\\+
\left.i\Omega_RT_1\sin2\theta\rho_-(t)\right]+I_Z\Gamma_-.
\end{multline}
This equation can be rewritten as follows:
\begin{multline}\label{dotcurr1}
\frac{{d\langle\widehat{I}_q\rangle}}{{dt}}=-\frac{1}{T_1}\langle\widehat{I}_q\rangle
-iI_Z\Omega_R\sin2\theta\rho_-(t)+I_Z\Gamma_-.
\end{multline}
With the aid of the steady state solutions (\ref{st0}) -
(\ref{st-}) we find the stationary current when the interaction
between the qubit and the tank is absent:
\begin{equation}\label{st_curr}
    \langle\widehat{I}_q\rangle^{st}=-I_Z\rho^{(eq)}\frac{1+\delta^2T_2^2}{1+\delta^2T_2^2+T_1T_2\Omega_1^2},
\end{equation}
where $T_2=1/\Gamma_{\varphi}$.

It is not unexpected that this expression is quite similar to the
one for the longitudinal magnetization in NMR\cite{Bloch1}.
However, the important difference is that the frequency $\Omega_1$
depends essentially on the current components of the qubit
($\Omega_1=F\sqrt{I_X^2+I_Y^2}$).

For the Fourier component of
(\ref{dotcurr})we obtain:
\begin{multline}\label{currFourier}
\left(\frac{{d\langle\widehat{I}_q\rangle}}{{dt}}\right)_{\omega}=
-\frac{2\pi M}{\Phi_0}I_T(\omega)\frac{I_Z
}{T_1}\left[\chi_{\rho}(\omega)\cos2\theta\right.\\+
\left.\chi_{\rho_+}(\omega)\sin2\theta+
i\Omega_RT_1\sin2\theta\chi_{\rho_-}(\omega)\right].
\end{multline}
Therefore, for irradiated qubit the quantities $Z_1(\omega)$ and
$Z_2(\omega)$ are as follows:
\begin{multline}\label{Z1_irr}
Z_1(\omega)=-\frac{2\pi}{\Phi_0}\frac{I_Z
}{T_1}\left[\chi'_{\rho}(\omega)\cos2\theta\right.\\+
\left.\chi'_{\rho_+}(\omega)\sin2\theta
-\Omega_RT_1\chi''_{\rho_-}(\omega)\sin2\theta\right],
\end{multline}
\begin{multline}\label{Z2_irr}
Z_2(\omega)=-\frac{2\pi}{\Phi_0}\frac{I_Z
}{T_1}\left[\chi''_{\rho}(\omega)\cos2\theta\right.\\+
\left.\chi''_{\rho_+}(\omega)\sin2\theta
+\Omega_RT_1\chi'_{\rho_-}(\omega)\sin2\theta\right].
\end{multline}
In this case the tank detuning $\xi$ and the friction $\Gamma_T$
can readily be obtained from (\ref{ksi}) and (\ref{GammaT}) by
using the expressions (\ref{Z2_irr}) and (\ref{Z1_irr}) for
$Z_2(\omega)$ and $Z_1(\omega)$, respectively.
\begin{multline}\label{ksi1}
    \xi=\omega_T^2-\omega^2+\rho^{(eq)}\frac{2k^2L_qI_Z^2\omega_T^2}{\omega\hbar\Omega_RT_1}
    \frac{\delta\Gamma_{\varphi}\Omega_1}{\Gamma_{\varphi}^2+\delta^2+T_1\Gamma_{\varphi}\Omega_1^2}\\\times
    \left[-f_{\rho}''(\omega)+\frac{\Omega_1}{\Omega_R}f_{\rho_+}''(\omega)
    -\frac{\Omega_RT_1\Omega_1}{\Gamma_{\varphi}}f_{\rho_-}'(\omega)\right],
\end{multline}
\begin{multline}\label{GammaT1}
    \Gamma_T=\gamma_T-\rho^{(eq)}\frac{2k^2L_qI_Z^2\omega_T^2}{\omega^2\hbar\Omega_RT_1}
    \frac{\delta\Gamma_{\varphi}\Omega_1}{\Gamma_{\varphi}^2+\delta^2+T_1\Gamma_{\varphi}\Omega_1^2}\\\times
    \left[-f_{\rho}'(\omega)+\frac{\Omega_1}{\Omega_R}f_{\rho_+}'(\omega)
    +\frac{\Omega_RT_1\Omega_1}{\Gamma_{\varphi}}f_{\rho_-}''(\omega)\right],
\end{multline}
where
\begin{multline}\label{f1}
    f_{\rho}(\omega)=\frac{1}{D(\omega)}\left[\sin2\theta
    \left[\left(i\omega+\Gamma_{\varphi}\right)\left(i\omega+\frac{1}{T_1}\right)+\Omega^2_R\right]\right.\\+
   \left.\frac{\Omega_R^2}{\Gamma_{\varphi}}B\cos2\theta\right],
\end{multline}
\begin{equation}\label{f}
  f_{\rho_+}(\omega)=\frac{1}{D(\omega)}\left[\left(i\omega+
  \Gamma_{\varphi}\right)\left(i\omega+\frac{1}{T_1}\right)
  -\left(i\omega+A_1\right)\frac{\Omega_R^2}{\Gamma_{\varphi}}\right],
\end{equation}
\begin{equation}\label{f3}
   f_{\rho_-}(\omega)=\frac{i}{D(\omega)}\left(i\omega+
   \frac{1}{T_1}\right)\left(i\omega+2\Gamma_{\varphi}\right).
\end{equation}

The expressions (\ref{ksi1}) and (\ref{GammaT1}) can equally be
used for the flux and the charge qubit with due account for the
different current components ((\ref{Curr_comp_flux}) for the flux
qubit and (\ref{A}) - (\ref{D}) for the charge qubit), and the
different structure of the energy gap ((\ref{engap}) for the flux
qubit and (\ref{En_ch_qb}) for the charge qubit).

\subsection{Damping and amplification by a superconducting flux qubit}

A remarkable property of the irradiated flux qubit coupled to a
low frequency $LC$ circuit is that the effective quality factor of
the tank can be made several times higher (lower) than that for an
unloaded tank. A clear phenomenological explanation of this
effect, which has been found experimentally, has been given in
Ref. \onlinecite{Graj}, where the effect has also been explained
by a rigorous quantum treatment of the problem followed by a
numerical solution of the corresponding equations for the global
density matrix of the qubit-tank system.

Here we show analytically that in the frame of our approach the
effect results from the frequency dependent quality factor of the
tank coupled to an irradiated qubit. The analytical expression we
obtain for the quality factor is valid for small high frequency
detuning, $\delta<<\omega_0$, $\Delta E/\hbar$ which allows us to
account only for the transitions between the Rabi levels. As is
seen from (\ref{GammaT1}) the amplification occurs if the high
frequency detuning $\delta$ is positive. It means that on the
energy diagram of the qubit the working point of the driving
current from the tank is located to the left from the point of the
high frequency resonance (see Fig.~1 in Ref.~\onlinecite{Graj}).
Positive values of $\delta$ correspond to the inverted population
of the Rabi levels (see Eq. (\ref{st0})). It is just the emission
of these Rabi photons which pumps the energy into the tank. For
negative values of $\delta$ the working point of the driving
current from the tank is located to the right from the point of
the high frequency resonance. For this case the lower Rabi level
is more populated than the higher level, and the tank transfers
the energy to the Rabi levels.

From (\ref{GammaT1}) the quality factor of the  tank coupled to
the qubit can be written as
$\frac{1}{Q}=\frac{1}{Q_T}+\frac{1}{Q_{qb}}$, where
\begin{multline}\label{Q_qb}
    \frac{1}{Q_{qb}}=-\rho^{(eq)}\frac{2k^2L_qI_q^2\varepsilon^2\omega_T}
    {\Delta_{\varepsilon}^2\omega^2\hbar\Omega_RT_1}
    \frac{\delta\Gamma_{\varphi}\Omega_1}{\Gamma_{\varphi}^2+\delta^2+T_1\Gamma_{\varphi}\Omega_1^2}\\\times
    \left[-f_{\rho}'(\omega)+\frac{\Omega_1}{\Omega_R}f_{\rho_+}'(\omega)
    +\frac{\Omega_RT_1\Omega_1}{\Gamma_{\varphi}}f_{\rho_-}''(\omega)\right].
\end{multline}
As is seen from (\ref{Q_qb}), $Q=Q_T$ at the point $f_X=0$
($\varepsilon=2\Phi_0I_q f_X$). The reason for this is the
vanishing of the current component $I_Z$ at this point (see
(\ref{Curr_comp_flux})).

We calculate from expression (\ref{GammaT1}) the frequency
dependence  of the tank quality factor $Q$ and the
voltage-frequency curve for the tank coupled to the irradiated
qubit. We take the following parameters of the tank, $Q_T=300$,
$k^2=0.001$, $\omega_T=2\pi\times 50$ MHz, and of the qubit,
$L_q=25$ pH, $I_q=300$ nA, $\Gamma_{\varphi}=4\times 10^7$
c$^{-1}$, $T_1=1.25\times 10^{-8}$ c, $\Delta/h=3.78$ GHz. For the
microwave frequency we take $\omega_0=2\pi\times 3.81$ GHz, and
the microwave power in the frequency units is $F/h=2\pi\times 45$
MHz.

\begin{figure}
  \includegraphics[width=6 cm, angle=-90]{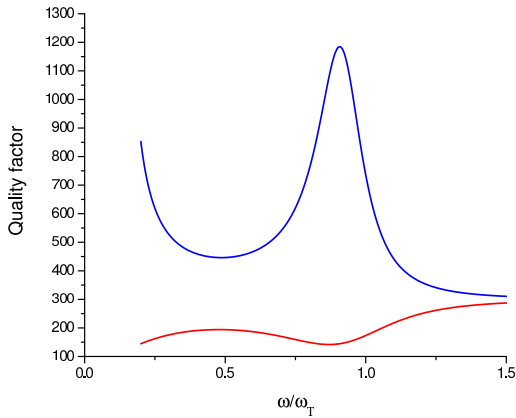}\\
\caption{(Color online). The dependence of the quality factor of
the tank coupled to an irradiated qubit. The quality of unloaded
tank $Q_T=300$. The upper curve ($f_X=0.0027, \delta>0$) shows the
effect of the amplification. The lower curve ($f_X=0.004,
\delta<0$) demonstrates the effect of the damping.}\label{Quality}
\end{figure}

The dependence of the tank quality factor on the low frequency is
shown in Fig.~\ref{Quality}.

\begin{figure}
  \includegraphics[width=6 cm, angle=-90]{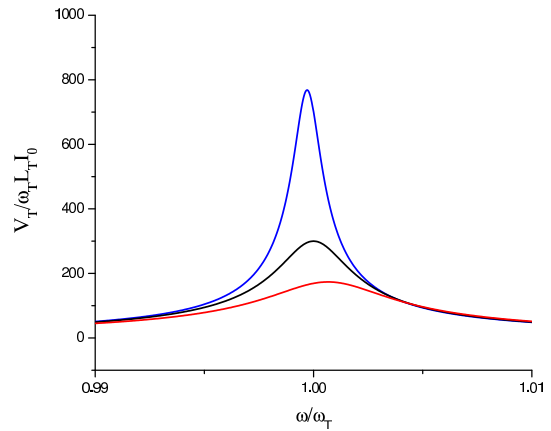}\\
\caption{(Color online). The dependence of the normalized tank
voltage on the low frequency. The quality factor of the unloaded
tank $Q_T=300$. The upper curve ($f_X=0.0027, \delta>0$) shows the
effect of the amplification. The lower curve ($f_X=0.004,
\delta<0$) is the effect of the damping. The curve, which is in
between, is the voltage curve for the unloaded
tank.}\label{Voltage}
\end{figure}

\begin{figure}
  \includegraphics[width=6 cm, angle=-90]{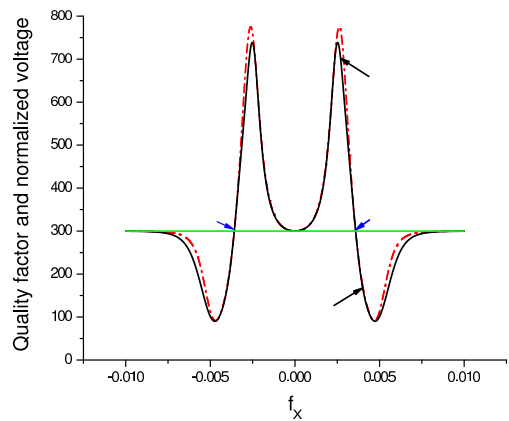}\\
\caption{(Color online). The dependence of the normalized tank
voltage (black curve) and the tank quality factor (red dash-dot
curve) on the bias flux $f_X$, plotted for $\omega=\omega_T$. The
quality factor of the unloaded tank ($Q_T=300$) is shown by the
green straight line. The black (long) arrows show the points which
correspond to the voltage at the tank resonance in
Fig.~\ref{Voltage}. The blue (short) arrows show the points where
the microwave frequency matches exactly the gap frequency
$\omega_0=\Delta_{\varepsilon}/\hbar$.}\label{Voltage_vs_Flux}
\end{figure}

The dependence of the normalized tank voltage on the low frequency
(bias frequency of the tank) calculated for the same data as those
for Fig.~\ref{Quality} is shown in Fig.~\ref{Voltage}. For the
same data we also show in Fig.~\ref{Voltage_vs_Flux} the
dependence of the voltage and the tank quality on the bias DC flux
$f_X$. The curves are plotted for a tank resonance
$\omega=\omega_T$. The blue (short) arrows at this figure show two
points where the microwave frequency matches exactly the gap
frequency $\omega_0=\Delta_{\varepsilon}/\hbar$. Between these
points the high frequency detuning $\delta$ is positive, and the
Rabi photons pump the energy into the tank. Beyond them the high
frequency detuning $\delta$ is negative, and the energy is drained
off the tank.

The analysis of Eqs. (\ref{ksi1}) and (\ref{GammaT1}) shows that
the detuning of the tank due to the interaction with the qubit is
rather small, therefore, the resonance frequency of the coupled
tank is practically equal to the unloaded value $\omega_T$.
However, the quality factor of the coupled tank has a strong
dependence on the frequency. Generally, the position of the peak
value of the quality factor does not coincide with the resonance
frequency of the tank (see Fig.~\ref{Quality}). Therefore, the
tank voltage curve can be appreciably differed from the lorentzian
line.
\section{Continuous readout of a superconducting qubit using a DC SQUID in the inductive mode}
\begin{figure}
  \includegraphics[width=6 cm, angle=-90]{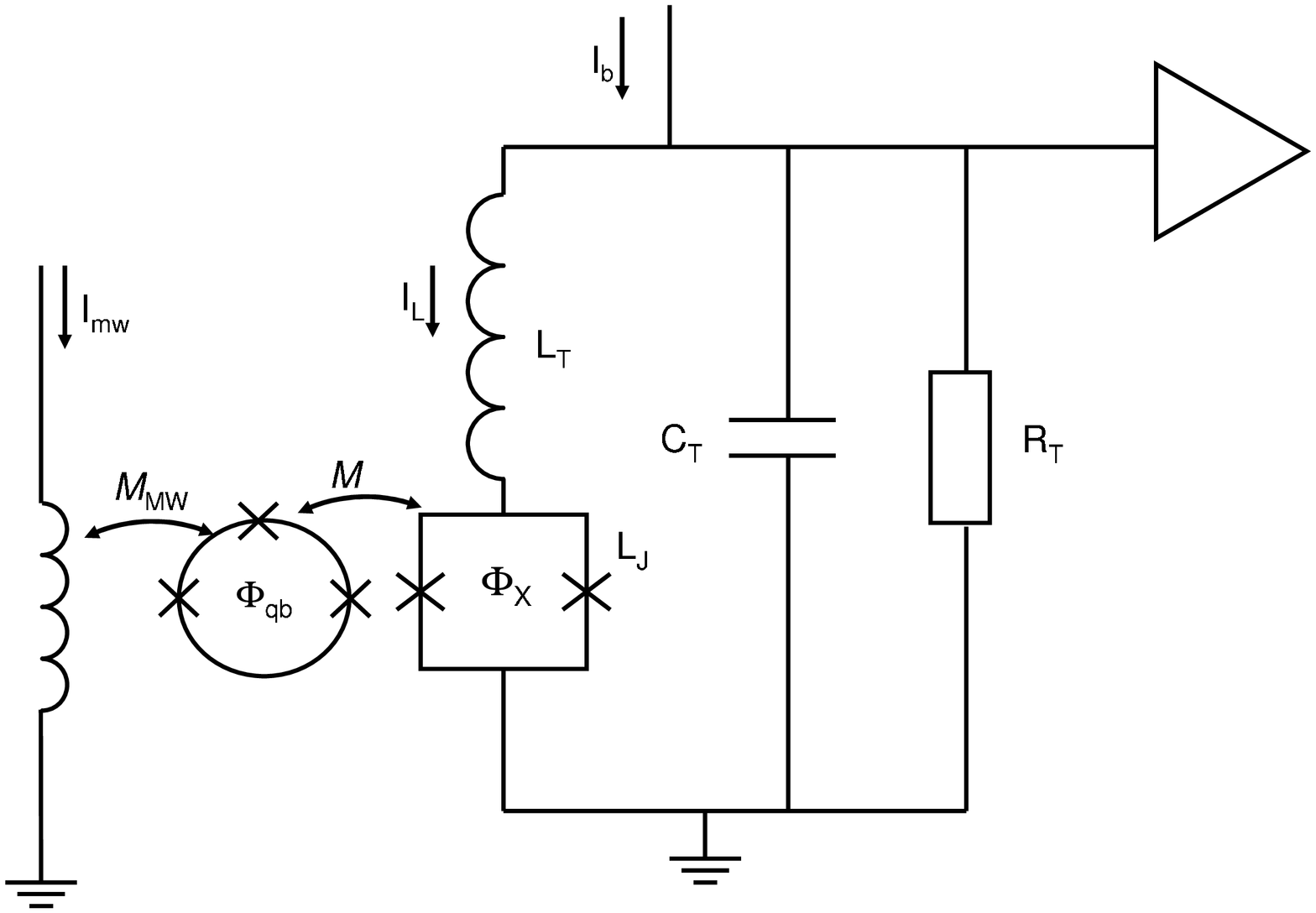}\\
\caption{Flux qubit coupled to the inductive DC SQUID, which is
incorporated in a resonant tank circuit.}\label{circuit}
\end{figure}

This method has been realized in Refs.~\onlinecite{Lup} and
\onlinecite{Lup2}. In these works a DC SQUID operating in the
inductive mode (the bias current through the SQUID is less than
its critical current) was inductively coupled to a flux qubit. To
enhance the measurement sensitivity, DC SQUID was incorporated in
a resonant $LC$ circuit (see Fig.~\ref{circuit}). The
spectroscopic measurements in Refs.~\onlinecite{Lup} and
\onlinecite{Lup2} have been performed by measuring the switching
currents of the DC SQUID after the microwave pulse had been
applied to the qubit.

Here we propose the modification of the method which does not
require a measurement of the switching currents. Our method is
based on the detection of the response of the resonant $LC$
circuit in which a DC SQUID is incorporated. It is assumed that in
the process of the measurement the qubit is irradiated by
microwaves continuously. We show that the DC SQUID in the
inductive mode directly measures the $I_Z$ component of the
current in a qubit loop. Our approach can be applied both for flux
and charge qubits.

First, we write the well known expression for the current across
the DC SQUID
\begin{equation}\label{dccurr}
    I_L=I_C\cos(f_{sq})\sin\gamma,
\end{equation}
where $I_L$ is the current in the inductor $L_T$, which is
connected in series with the SQUID loop, $I_C$ is the critical
current of the SQUID, $\gamma$ is the Josephson phase, which is
related to the voltage across a SQUID by the Josephson expression
$V_{sq}=\frac{\Phi_0}{2\pi}\frac{d\gamma}{dt}$. The quantity
$f_{sq}$ is the normalized total flux in the SQUID loop,
$\Phi_{sq}$, which is the sum of the external control flux
$\Phi_X$ and the flux $M\langle\widehat{I}_q\rangle$ from the
qubit:
\begin{equation}\label{flux}
    f_{sq}\equiv\frac{\pi\Phi_{sq}}{\Phi_0}=\frac{\pi\Phi_X}{\Phi_0}+
    \frac{\pi M\langle\widehat{I}_q\rangle}{\Phi_0}.
\end{equation}
In this expression we neglected the flux $MI_{circ}$,  which is
generated in the SQUID loop by a circulating current $I_{circ}$.
Below we assume $I_L<<I_C$, which implies $\gamma<<1$. From
(\ref{dccurr}) we get $\gamma=I_L/I_C\cos f_{sq}$, and for the
voltage across the tank we have:
\begin{equation}\label{volt_sq}
    V=\left(L_T+L_J\right)\frac{dI_L}{dt},
\end{equation}
where
\begin{equation}\label{ind_Jos}
    L_J=\frac{\Phi_0}{2\pi I_C\cos f_{sq}}
\end{equation}
is the Josephson inductance of the DC SQUID.
The equation for the voltage across the tank is similar to (\ref{volt}):
\begin{equation}\label{volt_sq_1}
     \ddot V + \gamma _T \dot V + \frac{\omega _T^2}{1+\frac{L_J}{L_T}} V =   \omega _T^2 L_T \dot
     I_b.
\end{equation}
We assume the coupling of the qubit to the tank is weak:
$\frac{\pi M\langle\widehat{I}_q\rangle}{\Phi_0}<<1$. For this
case Eq.~(\ref{volt_sq_1}) can be written in the following form:
\begin{multline}\label{volt_sq_approx}
     \ddot V + \gamma _T \dot V + \omega _T^2\frac{\cos f_X}{\cos f_X+\frac{L_J^{(0)}}{L_T}} V\\-
      \omega_T^2\frac{k}{2}\left(\frac{L_q}{L_T}\right)^
      \frac{1}{2}\frac{\sin f_X}{\left(\cos f_X+\frac{L_J^{(0)}}
      {L_T}\right)^2}\frac{\langle\widehat{I}_q\rangle}{I_C}V= \omega _T^2 L_T \dot I_b,
\end{multline}
where $f_X=\frac{\pi\Phi_X}{\Phi_0}$, $L_J^{(0)}=\frac{\Phi_0}{2\pi I_C}$.

As it is seen from this expression the tank response is
proportional to the averaged qubit current, but not to its time
derivative. Therefore, we may take for the quantity
$\langle\widehat{I}_q\rangle$ in (\ref{volt_sq_approx}) its
stationary time independent part
$\langle\widehat{I}_q\rangle^{st}$. By doing this we may neglect
the off diagonal components of the density matrix $\rho_+$, and
$\rho_-$ since their stationary parts are proportional to the
coupling constant $k$, and their contribution to the tank response
scales as $k^2$.

Therefore, the influence of the qubit results in the tank detuning
\begin{equation}\label{detun}
    \xi=\omega_T^2-\omega^2-\omega_T^2\frac{k}{\sqrt{2}}\left(\frac{L_q}{L_T}
    \right)^\frac{1}{2}\frac{\langle\widehat{I}_q\rangle^{st}}{I_C},
\end{equation}
where we take $f_X=\pi/4$ and assume $L_J^{(0)}<<L_T$.
The detuning can be detected from the phase of the output signal at resonance:
\begin{equation}\label{phase}
    \tan\alpha=-\frac{kQ_T}{\sqrt{2}}\left(\frac{L_q}{L_T}\right)^\frac{1}{2}
    \frac{\langle\widehat{I}_q\rangle^{st}}{I_C}.
\end{equation}
The expressions (\ref{detun}) and (\ref{phase})can equally be
applied both for ground state and spectroscopic measurements. For
the ground state measurements (the microwaves are absent) we have
from (\ref{Curr1}) $\langle \widehat{I}_q\rangle^{st}
=-I_Z\rho^{(eq)}$. For spectroscopic measurements of  an
irradiated qubit $\langle\widehat{I}_q\rangle^{st}$ is given in
(\ref{st_curr}).

In fact, DC SQUID measures not an absolute value of the flux
threading its loop,  but the change of the flux. Therefore, we may
subtract the ground state measurements from those for an
irradiated qubit to obtain a pure lorentzian line
\begin{multline}\label{Lorentz}
    \left(\tan\alpha\right)_{gr}-\left(\tan\alpha\right)_{irr}=
    \frac{kQ_T}{\sqrt{2}}\left(\frac{L_q}{L_T}\right)^\frac{1}{2}\frac{I_Z}{I_C}\rho^{(eq)}\\\times
    \frac{T_1T_2\Omega_1^2}{1+\delta^2T_2^2+T_1T_2\Omega_1^2},
\end{multline}
which allows for the direct measurement from its width of the dephasing time $T_2$\cite{com2}.

This technique is similar to that used in low field NMR, where a
longitudinal magnetization can be measured directly with the aid
of SQUID (see Ref.~\onlinecite{Green2} and references therein).

In conclusion, we developed here a quantum theory for the
calculation  of the low frequency linear susceptibility of the
interferometer type superconducting qubits. The obtained general
results are applied for the calculation of the tank detuning and
friction both for flux and charge qubit. For irradiated flux qubit
we obtained explicit expression for the tank quality factor which
allows us to calculate the recently found effect of the
amplification and the damping of the tank. We have also shown the
application of the theory to the continuous radio-frequency
monitoring of the qubit with the aid of DC SQUID in the inductive
mode. Our theory shows that radio-frequency method can also be
applied for the investigation of other types of quantum two level
structures provided the interaction between the tank and quantum
system is weak.
\acknowledgements

Ya. S. G. thanks M. Grajcar and S. Shevchenko for fruitful
discussions. This work was supported by the DFG under grant IL
150/1-1 and partly by the ESF under AQDJJ programme. E.I. thanks
also the EU for support through the RSFQubit and the EuroSQIP
projects.


\begin{thebibliography}{99}
\bibitem{com3} This is different from the case of qubit embedded
in a resonant microwave circuit, where the resonator together
with a qubit are treated quantum mechanically\cite{Blais}.
\bibitem{com} It is true that if the system is subjected to a strong external
driving the $\Gamma$'s rates and the corresponding rate equations can
be substantially modified \cite{Smirnov1,Hartmann}. However, even in
this case the most of experiments can adequately be explained  by
the Bloch equations with phenomenological $\Gamma$'s rates as fitting parameters (see, for example,\cite{Saito}).
\bibitem{com1} Frequently, the term "Rabi frequency" is associated
with the quantity $\Omega_1$. Here we call by this term the
quantity $\Omega_R$, the frequency with which the population
oscillates if the high frequency detuning $\delta$ is different
from zero.
\bibitem{com2} This curve can also be measured from the distribution of
the switching currents with the help of underdamped DC SQUID\cite{Wal}.
\end{thebibliography}
\end{document}